\documentclass[prl,aps,showpacs,twocolumn]{revtex4}
\usepackage{graphicx}
\usepackage{bm}
\usepackage{mathrsfs}
\usepackage{amsmath}
\renewcommand{\L}{\mathscr{L}}

\begin{document}

\title{Atomtronic circuits of diodes and transistors}

\author{R. A. Pepino, J. Cooper, D. Z. Anderson and M. J. Holland}
\affiliation{JILA, National Institute of Standards and Technology and
  Department of Physics, University of Colorado, Boulder, Colorado
  80309, USA}

\date{\today}

\begin{abstract}
  We illustrate that open quantum systems composed of neutral,
  ultracold atoms in one-dimensional optical lattices can exhibit
  behavior analogous to semiconductor electronic circuits. A
  correspondence is demonstrated for bosonic atoms, and the
  experimental requirements to realize these devices are
  established. The analysis follows from a derivation of a quantum
  master equation for this general class of open quantum systems.
\end{abstract}

                            
\maketitle

Atomtronics \cite{Seaman} focuses on ultracold atom analogs of
electronic circuits and devices. Previous work has established the
possibility of diode \cite{diode,Raizen} and transistor
\cite{Zoller,Zozula,spinField} behavior in atomic systems. We
calculate the device characteristics of these semiconductor-like
systems by construction of complete atomtronic circuits. Such circuits
contain the analog of electronic power supplies or batteries, and
the necessary device connections. This opens up the possibility for
more complex circuits in which atomtronic device components can be
cascaded. The approach we introduce develops a general theoretical
method for solving the dynamics of strongly-interacting, many-body,
{\em open} quantum systems.

A diode is in essence a device which exhibits an asymmetric response; applying
a potential gradient one way leads to a large current and the opposite way
leads to virtually no current. The atomtronic diode we present is novel
compared to some previously proposed devices \cite{diode, Raizen} in that it
does this without relying on an intrinsic irreversibility of the device itself
({\em e.g.}\/ the spontaneous emission of a photon or phonon). The atomtronic
transistor presented here contrasts with other proposals \cite{Zoller,Zozula},
in that it operates in steady-state, rather than only in an initial (transient)
regime.  It also reproduces the essential transistor behavior, including the
control of a larger atomtronic current with a smaller one, the ability to
perform digital logic, and exhibits linear amplification. One motivation for
designing atomtronic diode and transistor components that are intrinsically
reversible is that they can then be combined in quantum computing applications
where coherent logic is required.

This topic is relevant for the emerging field of strongly-correlated
dynamical and nonequilibrium phenomena in ultracold atomic
gases~\cite{JZAnn,BlochRMP,BlochJPhysB,Lukin}. Atomtronic devices
require site-by-site control of the temporal or spatial properties of
optical lattices~\cite{OptLatticeRMP,greiner,Esslinger,Tonks}, a
common goal for numerous experiments aiming to utilize optical
lattices for quantum control, quantum transport, quantum computing,
and quantum simulation~\cite{BlochNat,Calarco,Chiara}.  What is novel
here is that the atomtronic systems we consider are open quantum
systems, meaning that they are composed of optical lattices coupled to
two or three reservoirs, with the reservoirs acting as sources or
sinks for particles.

The Hamiltonian takes the separable form
\begin{equation}
\hat{H}=\hat{H}_{\rm sys}+\hat{H}_{\rm res}+\hat{H}_V ,
\end{equation}
where $\hat{H}_{\rm sys}$ and $\hat{H}_{\rm res}$ denote system and reservoir
Hamiltonians, respectively, and $\hat{H}_V$ their interaction.
Although atomtronic devices can utilize either fermionic or bosonic
atoms, we focus our discussion on the bosonic case, providing contrast
with electronics where the carriers are fermionic. In the lowest band
approximation, $\hat{H}_{\rm sys}$ is the Bose-Hubbard Hamiltonian
that describes ultracold bosons in a one-dimensional optical lattice
\cite{greiner,jaksch,phillips}, given by
\begin{eqnarray}
  \hat{H}_{\rm sys} &\! =\! &{}-\sum_{\left\langle i,j\right\rangle }J_{ij}
  \hat{a}_{i}^{^{\dag
    }}\hat{a}_{j}+\sum_{i}\bigl( \epsilon
  _{i}\hat{N}_{i}+\frac{U}{2}\hat{%
    N}_{i}( \hat{N}_{i}-1) \bigr) ,
\label{eqn:BH}
\end{eqnarray}
where site $i$ has energy $\epsilon _{i}$, interaction energy $U$,
annihilation and number operators $\hat{a}_{i}$ and $\hat{N}_{i}=
\hat{a}_{i}^{\dag }\hat{a}_{i}$ respectively, and $J_{ij}$~is the
hopping energy between adjacent sites $i$ and~$j$. To achieve a
steady-state current, atomtronic circuits require two or more
reservoirs held at different chemical potentials. The free Hamiltonian
for the reservoirs is
\begin{equation}
  \hat{H}_{\rm res}=\sum_{\nu, l}\hbar\omega_{\nu l}
  \hat{R}_{\nu l}^\dag\hat{R}_{\nu l}\ ,
\end{equation}
where $\nu$ identifies the mode of reservoir $l$ with energy
$\hbar\omega_{\nu l}$ and annihilation operator $\hat{R}_{\nu l}$. It
is assumed that each reservoir is so large that its thermodynamic
properties are parametrized by a constant chemical potential $\mu_l$
and temperature $T_l$. Each reservoir is connected to a single system
site~$s_l$ so that the interaction between the system and reservoir
can be written as
\begin{equation}
  \hat{H}_V=\sum_{\nu, l} g_{\nu l}\hat{R}^\dag_{\nu l}
  \hat{a}_{s_l}+{\rm h.c.} ,
\label{PotL}
\end{equation}
where $g_{\nu l}$ are the system-reservoir hopping energies.

\begin{figure}
\begin{center}
\includegraphics[width=7cm]{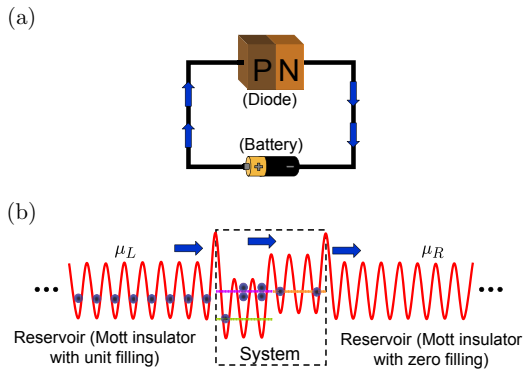}
\end{center}
\caption{(Color online.) (a) An electronic circuit with a battery,
  resistive wires, and a p-n junction diode in forward bias, and (b)
  the atomtronic equivalent. The left reservoir has chemical potential
  $\mu_L$ and unit filling, and supplies atoms that flow through the
  system (dashed box) to the right reservoir, with chemical potential
  $\mu_R$ and zero filling. The weak system-reservoir coupling is
  ensured by large tunneling barriers at each interface.}
\label{Fig1}
\end{figure}

Our atomtronic analog of electronic circuits begins with the
definition of the atomtronic battery, composed of two reservoirs of
ultracold atoms having different chemical potentials, corresponding to
different electric potentials at the terminals of a conventional
battery.  In Fig.~\ref{Fig1}(a), we show an electronic circuit
composed of a diode connected to a battery by resistive wires. In
Fig.~\ref{Fig1}(b), the atomtronic equivalent of this 
diode is the part of the lattice that is enclosed by a
dashed box with the lattice site energy on the left half offset from
the lattice site energy on the right half. To the left and right of
the diode are the reservoirs at high and low chemical potentials
respectively. These are the analogs of the battery terminals. The
system-reservoir separation is provided by large tunneling barriers
that ensure weak coupling. The explicit physical implementation for
the reservoirs shown is an optical lattice in the zero-temperature
Mott-insulator phase. In this system, strong interactions lead to a
fermionization of the bosons giving a broad continuum of reservoir
modes fully occupied below the chemical potential.

In the situation described we find that the current response depends
strongly on the direction of the applied chemical potential
difference.  The atomtronic diode's conduction asymmetry becomes
optimal when the height of the potential step in the middle of the
diode is equal to the on-site interaction energy $U$. This general
behavior is observed not only for the six-site lattice shown, but also for
lattices containing variable numbers of sites.
In order to explain the general mechanism, we consider the simplest
case of a two-well system with potential offset $U$ and connected on
each side to a reservoir. An appropriate basis for this two-site case
is the number basis $|n,m\rangle$, where $n$ and $m$ are the numbers
of atoms in the left and right wells respectively.

\begin{figure}
\begin{center}
\includegraphics[width=8cm]{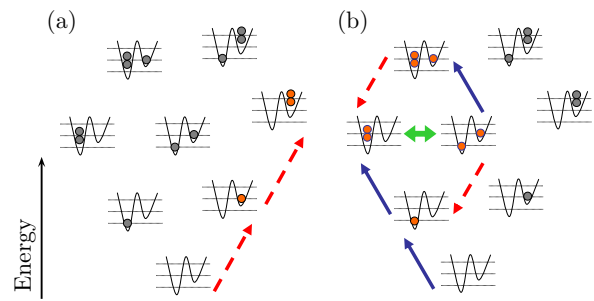}
\end{center}
\caption{(Color online.) Dynamical path of an initially empty two-well
  lattice system. The vertical axis corresponds to the energy of the
  system's quantum states.  The solid (blue) and dashed (red) arrows
  indicate the transfer of atoms from the left and right reservoirs
  respectively. (a) Reverse bias (low left hand chemical potential and
  high right hand chemical potential). Here the system evolves into
  the $|0,2\rangle$ dark state where it becomes trapped. (b) Forward
  bias (high left hand chemical potential and low right hand chemical
  potential).  The resonance between $|2,0\rangle$ and $|1,1\rangle$,
  shown by the double-headed arrow (green), allows the resonant
  transport of atoms across the lattice.}
\label{Fig2} 
\end{figure}

As shown pictorially in Fig.~\ref{Fig2}, the potential offset sets up
a degeneracy between two states.  The state $|2,0\rangle$ is resonantly
coupled to the state $|1,1\rangle$ (double headed arrow), and both
remain separated in energy from $|0,2\rangle$. The reverse bias
situation is shown in Fig.~\ref{Fig2}(a). If the chemical potentials
for both the left reservoir, $\mu_L$, and right reservoir, $\mu_R$,
are initially sufficiently low, the system begins in an equilibrium
devoid of particles. Raising $\mu_R$ from this point leads to
populating the $|0,1\rangle$ state. Since this state is detuned from
$|1,0\rangle$ by the on-site interaction energy $U$, the steady state
dynamics corresponds to virtually no current flow. Further raising
$\mu_R$ allows the state $|0,2\rangle$ to be occupied.  This state is
detuned from $|1,1\rangle$ by $2U$. As a result, there is again no
effective transport across the lattice system.

However, if $\mu_R$ is reset to its initial low value and $\mu_L$
raised, the resulting forward bias dynamics are quite different. As
shown in Fig.~\ref{Fig2}(b), when $\mu_L$ is increased to the point
where the $|2,0\rangle$ state is occupied, this opens up the
possibility for resonant transport of atoms from the left reservoir to
the right reservoir via the two resonant current loops;
$|2,0\rangle\!\rightarrow\!|1,1\rangle\!\rightarrow\!
|1,0\rangle\!\rightarrow\!|2,0\rangle$ and
$|2,0\rangle\!\rightarrow\!|1,1\rangle\!\rightarrow\!|
2,1\rangle\!\rightarrow\!|2,0\rangle$.

This qualitative picture of the diode operation is supported in detail
by the solution of the quantum dynamics for the open quantum
system. We solve for the evolution of the reduced density operator of
the system, $\hat{\sigma}\equiv {\rm Tr}_{\rm res}[\hat{\rho}]$, in which
$\hat{\rho}$ is the full density operator of the system and reservoir,
and the trace is over the reservoir degrees of freedom. Under the
Born-Markov approximation, the evolution of $\hat{\sigma}$ is given
by the master equation~\cite{Meistre,Cohen,Zubarev},
\begin{equation}
  \left(\frac{d}{dt}+i\hat{\L}_{\rm sys}\right)\hat{\sigma}(t)=
  \int_0^{\infty}d\tau\,\hat{{\mathscr R}}(\tau)\hat{\sigma}(t),
\label{Meqn}
\end{equation}
where the memory kernel is
\begin{equation}
  \hat{\mathscr R}(\tau)={\rm Tr}_{\rm res}\Bigl[i\hat{\L}_V 
  \exp[ -i\tau(\hat{\L}_{\rm sys}+\hat{\L}_{\rm res})]i\hat{\L}_V\Bigr]  .
\label{memkern}
\end{equation}
Here the Liouville operators $\hat{\L}_{\rm sys}$, $\hat{\L}_{\rm res}$, and
$\hat{\L}_V$ are defined via commutators of their respective
Hamiltonians, {\em i.e.\/} for operator $\hat{\Theta}$,
$\hbar\hat{\L}_X\hat{\Theta}=[\hat{H}_X,\hat{\Theta}]$,
$X\in\{{\rm sys},{\rm res},V\}$.

Substituting Eq.~(\ref{memkern}) into Eq.~(\ref{Meqn}) yields the
following closed form that is sufficiently simple to solve
numerically:
\begin{widetext}\begin{eqnarray}
    \frac{d\hat{\sigma}}{dt}-\frac{1}{i\hbar}
    \left[\hat{H}_{\rm sys},\hat{\sigma}\right] &=&
    -\frac{1}{\hbar^2}\sum_l |g_l|^2\sum_{\nu}
    \int_0^\infty d\tau \,e^{-\eta\tau}
    \bigg\{
    [\hat{a}_{s_l},\hat{a}_{s_l}^\dagger(-\tau)\hat\sigma]\langle
    \hat{R}^\dagger_{\nu l}\hat{R}_{\nu l}(-\tau)\rangle
    -[\hat{a}_{s_l},\hat\sigma\hat{a}_{s_l}^\dagger(-\tau)]\langle
    \hat{R}_{\nu l}(-\tau)\hat{R}^\dagger_{\nu l}\rangle\nonumber\\&&\quad
    +[\hat{a}^\dagger_{s_l},\hat{a}_{s_l}(-\tau)\hat\sigma]\langle
    \hat{R}_{\nu l}\hat{R}^\dagger_{\nu l}(-\tau)\rangle
    -[\hat{a}^\dagger_{s_l},\hat\sigma\hat{a}_{s_l}(-\tau)]\langle
    \hat{R}^\dagger_{\nu l}(-\tau)\hat{R}_{\nu l}\rangle\bigg\}
\label{masterEquation}
\end{eqnarray}\end{widetext}
with $\hat{a}_{s_l}(-\tau)={\rm exp}(-i\hat{H}_{\rm
  sys}\tau)\hat{a}_{s_l}{\rm exp}(i\hat{H}_{\rm sys}\tau)$ and
$\hat{R}_{\nu l}(-\tau)={\rm exp}(-i\hat{H}_{\rm res}\tau)\hat{R}_{\nu
  l}{\rm exp}(i\hat{H}_{\rm res}\tau)$ defining the system and
reservoir operators respectively in the interaction picture. The
parameter $\eta$ regularizes the integral and remedies the infrared
divergence that is needed to account correctly for the beyond
Born-Markov effects that arise from the coupling of near degenerate
system-reservoir levels~\cite{Satayan2}. The theory should be cutoff
and renormalized to account properly for the ultraviolet divergence
that arises from the replacement in Eq.~(\ref{masterEquation}) of
$g_{\nu l}$ by one that is mode independent, {\em i.e.}\/ $g_{\nu
  l}=g_l$.

\begin{figure}
\begin{center}
  \includegraphics[width=7cm,clip=true]{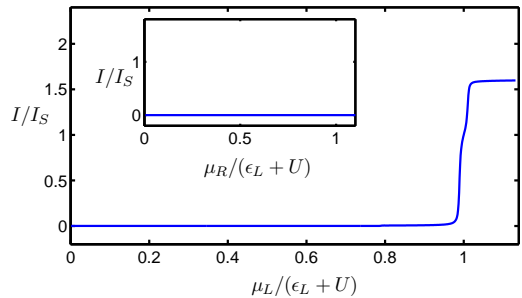}
\end{center}
\caption{(Color online.) Current of the forward-bias and reverse-bias
  (inset) response for the diode.  The horizontal axes are the
  chemical potential difference normalized to the resonance
  value. The vertical axes are the resulting current responses
  normalized to the saturation current $I_S$ that arises when the
  normalized chemical potential difference is unity.}
\label{Fig3}
\end{figure}

The numerical solution of the diode connected to a reservoir of this
type in steady state is shown in Fig.~\ref{Fig3}. The reverse bias
(inset) shows virtually no current response as a
function of the chemical potential difference, and the forward bias
case exhibits a turn-on of the current at the point where the
resonance states previously discussed become occupied. As anticipated
from the qualitative picture given earlier, when the current turns on,
there are two observable steps associated with linear combinations of
the two possible resonant current loops, with a rise broadened by
$\eta$.

\begin{figure}
\begin{center}
\includegraphics[width=7cm,clip=true]{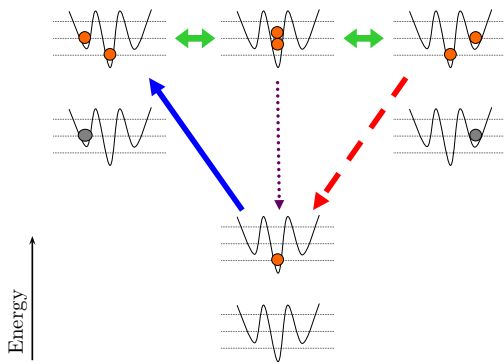}
\end{center}
\caption{(Color online.) The level scheme for the atomtronic
  transistor with a fixed chemical potential difference across the
  system. The base potential is set to put one particle on the middle
  site, which triggers a resonance that allows for transport across
  the system.  The left, right, and middle reservoir dynamics are
  represented by solid (blue), dashed (red), and dotted (purple),
  arrows respectively, while intrasystem dynamics are represented by
  double-headed arrows (green).}
\label{Fig4}
\end{figure}

Building on this understanding of the diode we now develop the
atomtronic transistor. Since a transistor is a three terminal device,
transistor action requires at least three lattice sites connected to
three independent reservoirs. As seen in Fig.~\ref{Fig4}, the middle
well is shifted down in energy by $U$ with respect to both the left
and right wells.  To investigate the device characteristics, we put a
fixed bias across the transistor. This means that $\mu_L$ is higher
than the value necessary to give an occupancy of one atom on the left
system site, and $\mu_R$ is below the value needed to deplete all
atoms on the right system site. We then study the system's
steady-state response to an increase of the middle chemical potential,
$\mu_M$.

\begin{figure}
\begin{center}
  \includegraphics[width=7cm,clip=true]{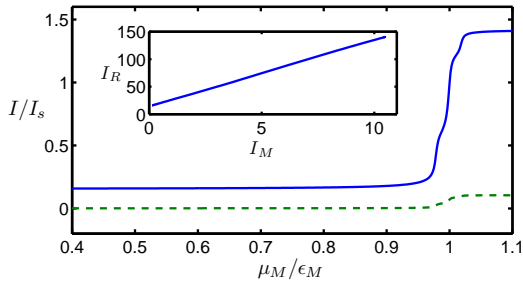}
\end{center}
\caption{(Color online). The current response and differential gain of
  the atomtronic transistor are plotted.  In this simulation, the
  reservoir couplings $g_l$ for the left and right reservoirs are five
  times that of the middle reservoir. For fixed $\mu_L-\mu_R$, we vary
  $\mu_M$ and record the response of currents leaving the system from
  the right site (blue solid curve) and from the middle site (green
  dashed curve).  Here, $I_S$ is the current value of the emitter when
  $\mu_M=\epsilon_M$.  The inset is a plot of right current versus the
  middle current, from which a large linear differential gain is
  apparent.}
\label{Fig5}
\end{figure}

When $\mu_M$ is low, the middle site contains virtually no atoms and
the system is locked in the $|1,0,0\rangle$. This  state is only
weakly coupled to $|0,0,1\rangle$ via a second-order off-resonant
process through $|0,1,0\rangle$.  As the middle chemical potential is
raised, a non-zero occupancy results on the middle site, and a
resonance is accessed among the states $|1,1,0\rangle$,
$|0,2,0\rangle$ and $|0,1,1\rangle$.  This triggers a cycle that
corresponds to a significant current across the system, as shown in
Fig.~\ref{Fig4}.  High linear differential gain (inset) is achieved by
ensuring that the system-reservoir coupling $g_l$ in
Eq.~(\ref{masterEquation}) is smaller in magnitude for the middle
reservoir compared to the left and right couplings. The current that
results from the $|0,2,0\rangle$ to $|0,1,0\rangle$ transition, {\em
  i.e.\/} an atom leaving the system and entering the middle
reservoir, is then much smaller than the current which flows from left
to right across the device. In Fig.~\ref{Fig5}, results are plotted of
the numerical simulation for the case in which the middle reservoir
coupling is one-fifth that of the reservoirs on either end.

In conclusion, we have calculated device characteristics of simple
diode and transistor atomtronic circuits. We have focused here on the
two-site and three-site models to simplify the presentation, but full
calculations for more extended systems with a greater site-number
confirm that the diode-like and transistor-like behavior
persists. This work motivates expansion of the analogy to more complex
circuits, such as amplifiers, constant current sources, flip-flops,
and logic elements. Apart from their potential practical value, it
would be intriguing to consider the connection of atomtronics with
quantum information physics and quantum computation, since computers
are based on cascaded logic devices and these atom devices presented
operate exclusively in the coherent regime.

Atomtronic devices will always be much slower than their electronic
counterparts.  For example, the overall current flow is determined
principally by the tunneling rate from one site to the
next~\cite{Seaman}. Compensating for this is the fact that atomtronic
studies are pursuing device physics in a novel physical system: unlike
electrons, atoms have a complex internal structure and internal
states, the atoms can be bosons or fermions, they are massive and
affected by gravitational and electromagnetic fields, the material
itself which is the optical lattice can be dynamically and spatially
varied, and many unavoidable effects present in solid-state systems
such as crystalline impurities, dislocations, and phonon scattering
are absent.

We would like to thank Rajiv Bhat, Meret Kr\"{a}mer, Dominic Meiser,
Brandon Peden, Brian Seaman, Jochen Wachter, and Christopher R. Pepino for their helpful
discussions.  We gratefully acknowledge support from the Airforce
Office of Scientific Research, the DOE, and the NSF PFC.

\end{document}